\documentclass[12pt]{spieman}  
\usepackage{amsmath,amsfonts,amssymb}
\usepackage{graphicx}
\usepackage{setspace}
\usepackage{tocloft}
\usepackage{hyperref}  
\usepackage{array} 
\usepackage{multirow} 
\usepackage[table]{xcolor}
\usepackage{colortbl}
\usepackage[left]{lineno} 
\usepackage{geometry}

\title{Observation of Hong-Ou-Mandel interference between photon and polariton}
\author[a,b,c\dag]{Yun-Ru Fan}
\author[a,b,c\dag]{Ying-Ao Su}
\author[d,$\ddagger$]{Kai Guo}
\author[e]{Bo-Yu Fan}
\author[a,b,c]{Yao-Qing Zhang}
\author[a,f]{Hai-Zhi Song}
\author[g]{Hao Li}
\author[e]{Yong Geng}
\author[b,h]{Kun Chen}
\author[b]{Deng-Ke Zhang}
\author[g]{Li-Xing You}
\author[a]{Yan-Yu Wei}
\author[a,b,c,i]{Guang-Can Guo}
\author[a,b,c,i,*]{Qiang Zhou}

\affil[a]{Institute of Fundamental and Frontier Sciences, University of Electronic Science and Technology of China, Chengdu 611731, China}
\affil[b]{Center for Quantum Internet, Tianfu Jiangxi Laboratory, Chengdu 641419, China}
\affil[c]{Key Laboratory of Quantum Physics and Photonic Quantum Information, Ministry of Education, University of Electronic Science and Technology of China, Chengdu 611731, China}

\affil[d]{Institute of Systems Engineering, AMS, Beijing 100141, China}

\affil[e]{Key Lab of Optical Fiber Sensing and Communication Networks, University of Electronic Science and Technology of China, Chengdu 611731, China}

\affil[f]{Southwest Institute of Technical Physics, Chengdu 610041, China}
\affil[g]{ Shanghai Institute of Microsystem and Information Technology, Chinese Academy of Sciences, Shanghai 200050, China}

\affil[h]{School of Optoelectronic Science and Engineering, University of Electronic Science and Technology of China, Chengdu 611731, China}

\affil[i]{CAS Key Laboratory of Quantum Information, University of Science and Technology of China, Hefei 230026, China}

\affil[$\dag$]{These authors contributed equally to this work.}

\cftpagenumbersoff{figure}
\cftpagenumbersoff{table} 
\begin{document} 
\maketitle

\begin{abstract}
Light–matter interactions underlie many quantum technologies, yet whether quasiparticles formed from such interactions preserve the full quantum state of light remains unresolved. Surface plasmon polaritons (SPPs), a class of polaritons formed by interacting photons with free-electron oscillations at metal-dielectric interfaces, are prime candidates to explore this question. Here we demonstrate quantum interference between single photons and SPPs using an Au–SiN\textsubscript{x} integrated photonic-plasmonic device. Our results reveal that SPPs retain the indistinguishability of their excitation photons, establishing SPP as a viable quantum information carrier and opening a potential route toward photonic–plasmonic quantum circuitry.
\end{abstract}

\keywords{surface plasmon polariton, photon--polariton interference, light--matter interaction, integrated quantum photonics}

{\noindent \footnotesize\textbf{$\ddagger$} Kai Guo, \linkable{guokai07203@hotmail.com} }
{\noindent \footnotesize\textbf{*} Qiang Zhou, \linkable{zhouqiang@uestc.edu.cn} }

\begin{spacing}{1.5}   

\section{Introduction}
\label{sect1}  
\begin{figure}[!h]
    \centering
    \includegraphics[width=8 cm]{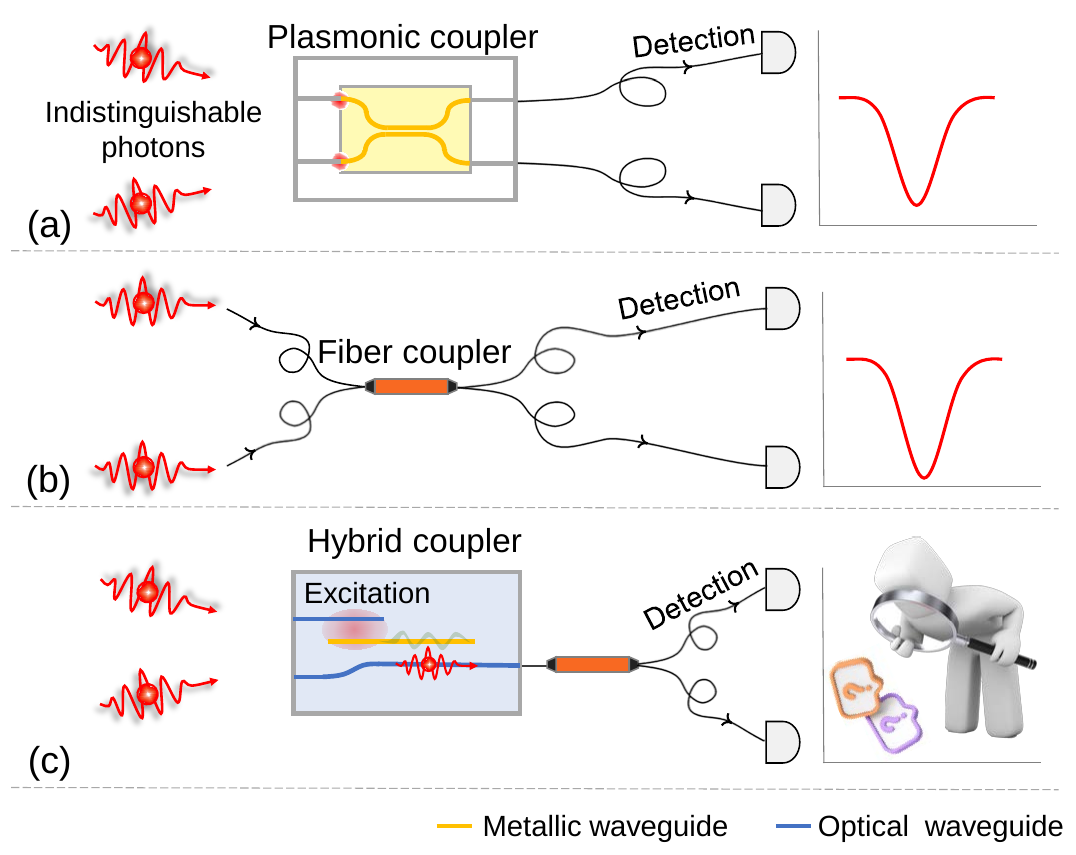}
    \caption{\textbf{HOM interference.} \textbf{a}, Polariton-polariton interference using a metallic directional coupler. \textbf{b}, Photon-photon interference using a fiber-based coupler. \textbf{c}, Proposed photon–polariton interference scheme using a hybrid device.}
    \label{fig:Fig1}
\end{figure}

Light–matter interactions form the cornerstone of many fundamental optical phenomena and modern photonic technologies \cite{boyd2008nonlinear,  flick2018strong}. Strong coupling between light and quantum excitations in a material gives rise to hybrid quasiparticles known as polaritons \cite{raether2006surface, butov2012behaviour, liu2015strong, frisk2019ultrastrong}, which simultaneously exhibit properties of photons and matter waves.~This hybrid nature enables polaritons to support field confinement and enhanced light–matter interactions \cite{nie1997probing, fang2005sub, anker2008biosensing, gramotnev2010plasmonics, yu2019plasmon, yu2022plasmon, 2024NatureRM}. Surface plasmon polariton (SPP) - formed by the coupling of photons with collective oscillations of free electrons at a metal–dielectric interface - is a paradigmatic example of a hybrid light–matter quasiparticle \cite{NATURE2003Surface, berini2009long, heeres2010chip, di2012quantum, berini2012surface, tame2013quantum}. The quantum nature of SPPs has been widely studied, with experimental confirmation of their wave–particle duality \cite{NP2009Wave-particle, NC2015Simultaneous} and preservation of key quantum properties such as bosonic behavior \cite{fujii2014direct, dheur2016single}, entanglement \cite{NATURE2002Plasmon-assisted, PRL2005Energy-time}, and indistinguishability \cite{fujii2012OL}. Among these, SPPs have been shown to exhibit the behavior of indistinguishable quantum quasiparticles \cite{NN2013Quantuminterference, PRA2014High-Visibility, NP2014Two-plasmon, PRA2014Observation}. In particular, Hong-Ou-Mandel (HOM) interference between two SPPs has been demonstrated using a plasmonic directional coupler \cite{NN2013Quantuminterference, NP2014Two-plasmon}. As illustrated in Fig.~\ref{fig:Fig1}a, two indistinguishable photons are incident on two separate input ports of a metallic directional coupler and excite SPPs. The SPPs propagate and interfere within the coupler with a 50:50 coupling ratio, leading to a pronounced HOM dip in the coincidence measurements. This observation confirms the second-order quantum interference of SPPs, analogous to the HOM interference of their excitation photons observed in a fiber-based setup, as shown in Fig.~\ref{fig:Fig1}b \cite{hong1987measurement, zou1991induced, lyons2018attosecond, fan2021effect, ndagano2022quantum, jin2024quantum}.

Despite these advances, previous demonstrations of HOM interference between two SPPs merely confirm that the SPPs themselves are mutually indistinguishable, i.e., SPPs can interfere as indistinguishable quantum quasiparticles. However, whether a single SPP fully preserves quantum states of its excitation photon, such as wavepacket structure and temporal coherence, remains an open question.~The question determines whether SPPs can serve not only as intermediaries for photon conversion, but also as true carriers of quantum information, which is essential for advancing our understanding of the basic property of polariton and the potential role of photonic–plasmonic system in future quantum technologies.

Here, we propose a photon–polariton quantum interference scheme and experimentally demonstrate the HOM interference using an Au–SiN\textsubscript{x} integrated photonic-plasmonic device, as shown in Fig.~\ref{fig:Fig1}c. By directly observing the quantum interference behavior of photons and SPPs, we reveal that the light–matter interaction process of SPP excitation preserves all properties of the input photons. Our findings provide new insight into the fundamental nature of such a light–matter interaction process, which bridges the gap between quantum optics and nanoplasmonics.

\section{Schematic design of the Au-SiN\textsubscript{x} integrated photonic-plasmonic device}
\begin{figure*}[!t]
    \centering
    \includegraphics[width=15 cm]{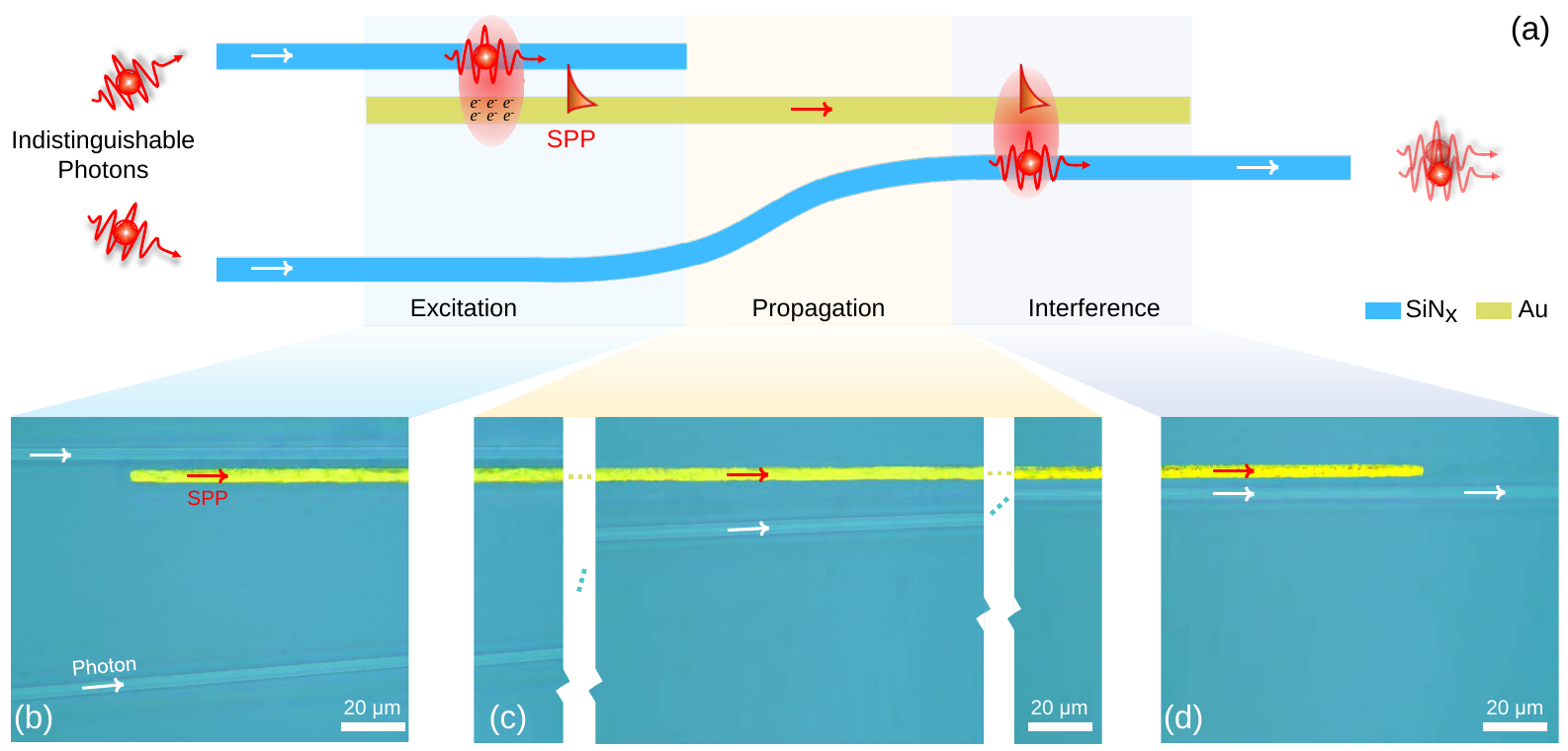}
    \caption{\textbf{Photon–polariton quantum interference device and experimental setup.} \textbf{a}, Schematic design of the Au-SiN\textsubscript{x} integrated photonic-plasmonic device with three functional regions: excitation, propagation, and interference. \textbf{b},\textbf{c},\textbf{d}, Optical microscope images of the excitation, propagation, and interference regions of the fabricated device, respectively.}
    \label{fig:Fig2}
\end{figure*}

Figure {\ref{fig:Fig2}}a illustrates the schematic design of the Au-SiN\textsubscript{x} integrated photonic-plasmonic device. It comprises two optical input ports, one output port, and three functional regions: excitation, propagation, and interference. In the excitation region, SPPs are excited by coupling photons from the upper SiN\textsubscript{x} waveguide into the adjacent Au waveguide. In the propagation region, SPPs and photons travel separately along the upper metallic and lower dielectric waveguides, respectively. The interference region is engineered as a 50:50 coupler, enabling quantum interference between the two modes.

Figure {\ref{fig:Fig2}}b,c,d presents the optical images of the fabricated photonic-plasmonic device, corresponding to the three regions, respectively. See more details in the Appendix B for the fabrication processes for the device. All waveguides are clad in SiO\textsubscript{2}. In our experiments, the SiN\textsubscript{x} waveguides are 66 nm thick and 4 $\mu$m wide, and the Au waveguide has a thickness of 36 nm and a width of 4 $\mu$m. See more details in the Appendix A for the simulation result of the design. In the excitation region, the Au waveguide length is 600 $\mu$m, with a vertical gap of 2.5 $\mu$m between the Au and upper SiN\textsubscript{x} waveguides. In the interference region, the coupling structure serves as an integrated coupler between adjacent waveguides through evanescent field interaction. The coupling ratio can be precisely tuned by varying the waveguide gap and coupling length. For the HOM interference experiment, we employ a coupler designed to achieve a 50:50 coupling ratio, with an Au waveguide length of 240 $\mu$m and a gap of 3 $\mu$m between the Au waveguide and the lower SiN\textsubscript{x} waveguide \cite{luo2023hybrid}. 

\section{Characterization of the Au-SiN\textsubscript{x} integrated photonic-plasmonic device}

\begin{figure*}[!t]
    \centering
    \includegraphics[width=15 cm]{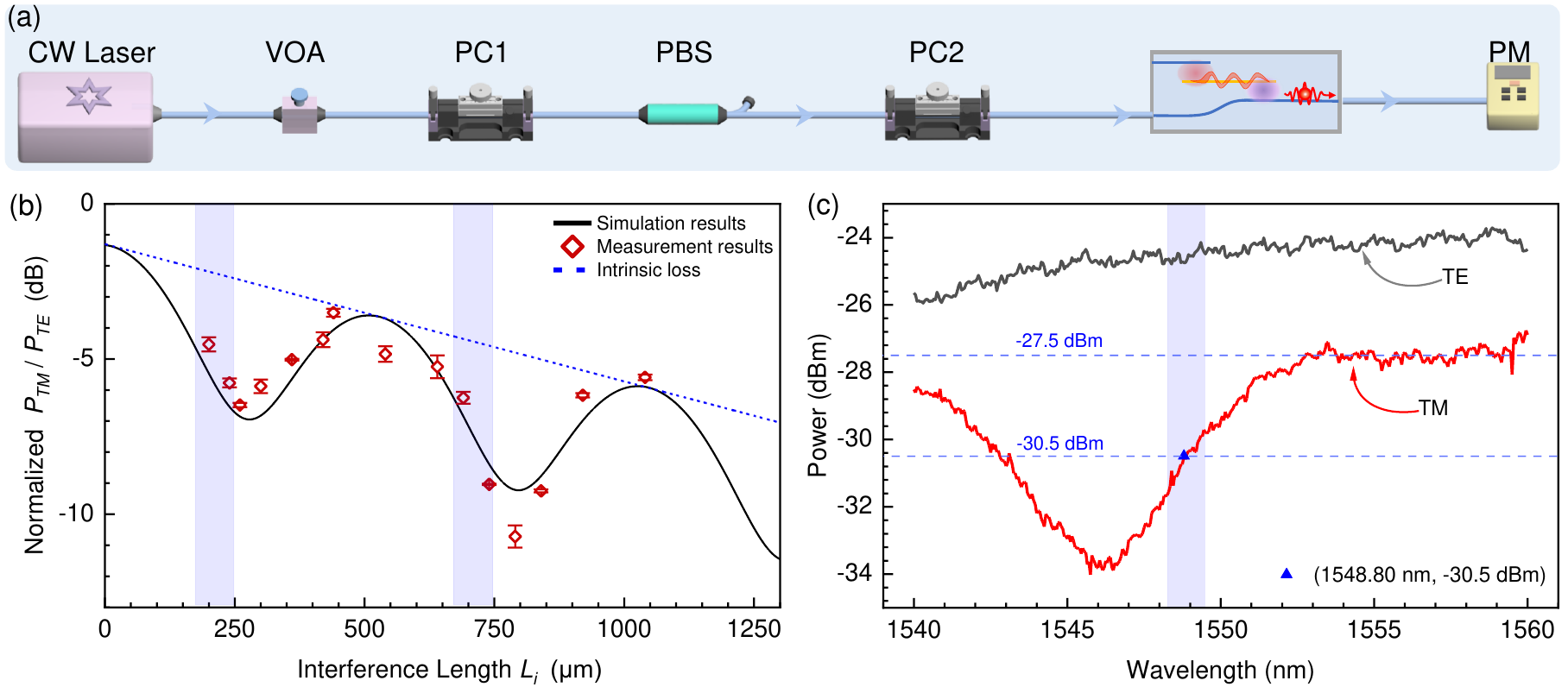}
    \caption{\textbf{Characterization of the fabricated device.} \textbf{a}, Transmission measurement of the integrated photonic-plasmonic device. \textbf{b}, Normalized P\textsubscript{TM}/P\textsubscript{TE} of SiN\textsubscript{x} waveguide with different coupling lengths. \textbf{c}, Output power of the SiN\textsubscript{x} waveguide at different input wavelengths at L\textsubscript{i} = 240 $\mu$m. VOA: variable optical attenuator; PC: polarization controller; PBS: polarization beam splitter; PM: power meter.}
    \label{fig:Fig3}
\end{figure*}

To characterize the fabricated integrated photonic-plasmonic device and determine the interference length that yields a 50:50 splitting ratio, we perform transmission measurements as shown in Fig.~\ref{fig:Fig3}a. A tunable laser (Toptica CTL pro), operating at a wavelength of 1549 nm with a linewidth of 10 kHz, serves as the light source. The power of the light is adjusted by a variable optical attenuator (VOA). Then the light passes through a polarization controller (PC1) and a polarization beam splitter (PBS), where PC1 aligns the input polarization to ensure linearly polarized light. A second polarization controller (PC2) is used to further adjust the polarization state to match the transverse magnetic (TM) or transverse electric (TE) mode of the on-chip waveguide. Light is coupled into the lower SiN\textsubscript{x} using a single-mode fiber. The transmitted power is recorded using an optical power meter (PM) by another single-mode fiber, with a total coupling loss of approximately 20 dB.

In the experiments, photons in the TE mode do not excite SPPs, the loss of which primarily originates from facet coupling and propagation in the SiN\textsubscript{x} waveguide. In contrast, photons in the TM mode experience additional losses due to SPP excitation, including mode conversion at the dielectric–metal interface, propagation loss in the Au waveguide, and scattering loss at the interference region. The output power exhibits a cosine-like dependence on the length \( L_{i} \), reflecting periodic energy exchange between the photonic and plasmonic modes. 

To characterize this behavior, we adjust the PC2 to obtain the maximum and minimum output power, corresponding to the TE and TM polarization states, respectively. To eliminate alignment-induced fluctuations and highlight the intrinsic mode-dependent loss, we normalize the power ratio as P\textsubscript{TM}/P\textsubscript{TE}. Figure~\ref{fig:Fig3}b shows the normalized ratio as a function of coupling length from 200 to 1040~$\mu$m. It can be seen that the power exhibits a cosine-like oscillation superimposed on a gradual attenuation trend, with a fitted propagation loss of approximately 1.94 dB/mm attributed to SPP excitation and propagation. 

The 50:50 splitting condition, defined as a 3 dB reduction in TM output power relative to TE, is achieved at coupling lengths of approximately 240~$\mu$m and 720~$\mu$m, indicating that nearly half of the optical power is successfully coupled into the Au waveguide and converted into SPPs, as shown by the shaded areas in Fig.~\ref{fig:Fig3}b. To minimize the influence of propagation loss in subsequent experiments, we select the shorter coupling length of 240~$\mu$m for HOM interference measurements. It is worth noting that discrepancy between the simulated and experimentally measured interference length \( L_{i} \) is primarily attributed to fabrication-induced variations, such as the thickness of the Au waveguide, the thickness of the SiN waveguide, and the gap between the Au and SiN\textsubscript{x} waveguides.

We further perform wavelength-dependent characterization of the photonic–plasmonic device at a fixed interference length of \( L_{i} \)=240~$\mu$m, as shown in Fig.~\ref{fig:Fig3}c. The tunable laser is swept across the 1540–1560 nm range with a step of 0.05 nm. For the TE mode, the output power remains relatively constant over the entire spectral range. In contrast, the TM mode exhibits a pronounced wavelength dependence, with a minimum in transmitted power near 1546 nm, corresponding to maximum coupling into the Au waveguide. For a coupling length of 240~$\mu$m, the wavelength range of 1548.3–1549.3 nm corresponds to a TM power drop of approximately 3 dB, indicating a splitting ratio varying from 40:60 to 60:40. Based on this characterization, the HOM interference measurements are performed at a central wavelength of 1549.02~nm, which falls within this optimal coupling window and closely approximates a 50:50 splitting condition — essential for achieving high-visibility two-photon interference.

\section{Schematic of the photon–polariton quantum interference with Au-SiN\textsubscript{x} integrated photonic-plasmonic device}
The schematic of our photon–polariton quantum interference is presented in Fig.~{\ref{fig:Fig1}}c. The state at the input of the interference region is written as:
\begin{equation}
\begin{aligned}
\left|\psi^{in}\right\rangle_{a b}=&\int d \omega_{1} \, \phi_{a}(\omega_{1}) \hat{a}^{\dagger}\left(\omega_{1}\right)\\  &\int d \omega_{2} \phi_{b}\left(\omega_{2}\right) \hat{b}^{\dagger}\left(\omega_{2}\right)  e^{-i \omega_{2} \tau}|0\rangle_{a b}
\end{aligned}
\end{equation}
where $\phi_{a, b}(\omega)$ denotes the spectral amplitude function of the input mode of SPP and photons, $\tau$ is the time delay between $a$ and $b$. The creation operators $\hat{a}^{\dagger}(\omega)$ and $\hat{b}^{\dagger}(\omega)$ satisfy bosonic commutation relations, and the states are normalized such that $\int d \omega|\phi(\omega)|^{2}=1$.

At the output of the interference region, the state takes the form 
\begin{equation}
\begin{aligned}
\left|\psi^{\text{out}}\right\rangle 
&= \frac{1}{4} \iint d\omega_1 d\omega_2\, \phi(\omega_1)\phi(\omega_2) e^{-i\omega_2 \tau} \\
&\quad \times \Big[
      \hat{e}^\dagger(\omega_1)\hat{e}^\dagger(\omega_2)
    + \hat{e}^\dagger(\omega_1)\hat{f}^\dagger(\omega_2)\\
  &  + \hat{f}^\dagger(\omega_1)\hat{e}^\dagger(\omega_2)
    + \hat{f}^\dagger(\omega_1)\hat{f}^\dagger(\omega_2)
\Big] |0\rangle.
\end{aligned}
\end{equation}
where $\hat{e}^{\dagger}(\omega)$ and $\hat{f}^{\dagger}(\omega)$ represents the creation operators of the output modes.

The component relevant to coincidence detection between outputs $e$ and $f$ is:
\begin{equation}
\begin{aligned}
\label{eq:s15}
|\psi_{\text{coinc}}(\tau)\rangle = &\frac{1}{4} \iint d\omega_1 d\omega_2\, \phi(\omega_1)\phi(\omega_2) e^{-i\omega_2 \tau}\\
&\times\left[ \hat{e}^\dagger(\omega_1)\hat{f}^\dagger(\omega_2) + \hat{f}^\dagger(\omega_1)\hat{e}^\dagger(\omega_2) \right] |0\rangle.
\end{aligned}
\end{equation}

The coincidence probability is given by:
\begin{equation}
\label{eq:4}
P(\tau) = \frac{1}{8} \left[ 1 + |g(\tau)|^2 \right]
\end{equation}
where $g(\tau)$ is the Fourier transform of the spectral intensity. For a Gaussian spectrum, Eq.~\ref{eq:4} reduces to 
\begin{equation}
P(\tau) = \frac{1}{8} \left[1 + e^{-2 \sigma^2 \tau^2} \right].
\end{equation}
In this case, the maximum and minimum coincidence probabilities are $p_{\max} = 1/4$ and $p_{\min} = 1/8$, respectively, yielding an ideal Hong–Ou–Mandel visibility of $V = (p_{\max} - p_{\min})/p_{\min} = 1$. Detailed derivations are provided in Appendix C.

\section{Photon–polariton quantum interference}
\begin{figure*}[!t]
    \centering
    \includegraphics[width=15 cm]{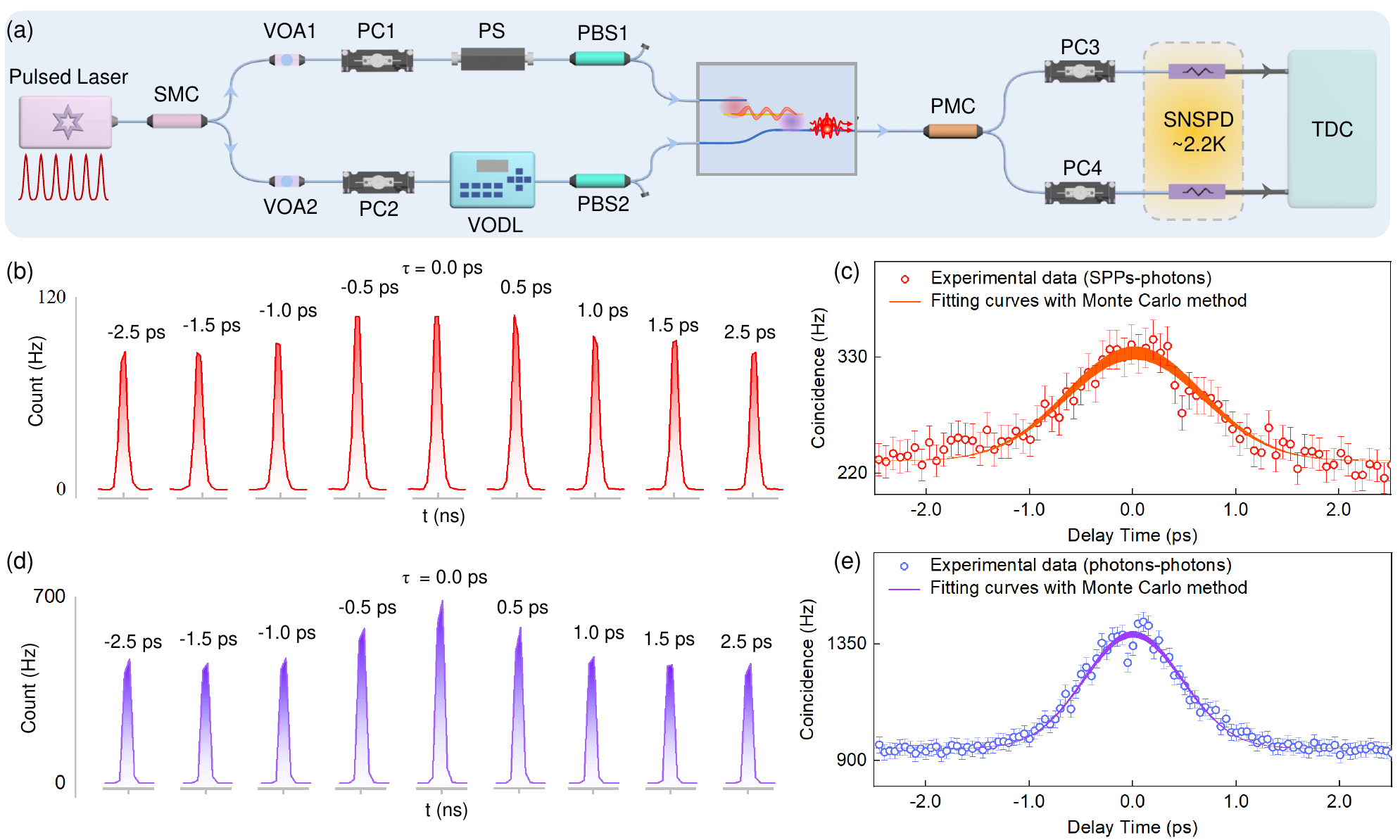}
    \caption{\textbf{HOM interference.} \textbf{a}, Experimental setup for HOM interference. \textbf{b}, Time-resolved coincidence histogram using the integrated photonic-plasmonic device. \textbf{c}, HOM interference peak showing indistinguishability between photons and SPPs. The FWHM is 1355.0$\pm$212.8 fs and the visibility is 47.87$\pm$4.32\%. \textbf{d}, Time-resolved coincidence histogram measured based on a 50:50 PMC. \textbf{e}, HOM interference curve obtained from the fiber-based setup with a visibility of 46.31$\pm$1.26\% and FWHM of 1143.8$\pm$37.9 fs. SMC: single-mode fiber coupler; VOA: variable optical attenuator; PC: polarization controller; PS: phase shifter; VODL: variable optical delay line; PBS: polarization beam splitter; PMC: polarization-maintaining fiber coupler; SNSPD: superconducting nanowire single photon detector; TDC: time-to-digital converter.}
    \label{fig:Fig4}
\end{figure*}
The experimental setup for HOM interference is illustrated in Fig.~\ref{fig:Fig4}a. A weak coherent-state single-photon source is realized by attenuating pulses from a mode-locked laser (PFL-200 M, Alnair Labs) centered at 1549.02 nm, with a spectral full width at half maximum (FWHM) of 3.52 nm. The laser operates at a repetition rate of 40 MHz with a measured pulse duration of 1.12$\pm$0.01~ps. A 50:50 single-mode fiber coupler (SMC) splits the laser pulses into two optical paths for HOM interference. The HOM interferometer consists of two variable optical attenuators (VOA1 and VOA2), two polarization controllers (PC1 and PC2), two polarization beam splitters (PBS1 and PBS2), a variable optical delay line (VODL), a phase shifter (PS), and the integrated photonic–plasmonic device. The VOA1 and VOA2 are used to attenuate the optical power to the single-photon level, while ensuring that the mean photon number is balanced between the two arms. Polarization indistinguishability between the two optical paths is achieved via the combination of PBSs and PCs. Temporal overlap of the two optical pulses is controlled using the VODL, which offers a tuning resolution of 1 fs and a total delay range of 560 ps. To suppress first-order interference, a PS is inserted into one arm to introduce phase shifts. See more details in Appendix D. The phase perturbations are driven by an arbitrary waveform generator (AWG), effectively randomizing the relative phase between the two arms. Then the photons exiting PBS1 and PBS2 are coupled into the integrated photonic-plasmonic device via a fiber array. Within the chip, SPPs are excited and interfere with photons propagating in the adjacent dielectric waveguide. The output photons from the chip are collected by the input port of a 50:50 polarization-maintaining fiber coupler (PMC). After passing through PCs, photons are detected by superconducting nanowire single-photon detectors (SNSPDs), which generate electrical pulses upon photon arrival time. These pulses are sent to a time-to-digital converter (TDC), which registers photon arrival times and identifies coincidence events. By scanning the relative delay between the two paths of the interferometer, we can observe a HOM interference signature as a characteristic peak in the coincidence counts, depending on the beam splitter configuration. 


We first characterize the indistinguishability of the SPPs and their excitation photons in the HOM experiment by performing a time-resolved correlation measurement\cite{PRA2014High-Visibility}. The recorded coincidence histograms at different relative time delays are shown in Fig.~\ref{fig:Fig4}b. When overlapping the photons and SPPs in time, the correlation peaks appear, where the wave-packets of photons and SPPs bunch. One half of the interference event travels to the same lower SiN\textsubscript{x} waveguide and output from the device, while the other half of the interference event is not coupled out from our integrated device. In Fig.~\ref{fig:Fig4}c, we plot the calculated area under the curves in Fig.~\ref{fig:Fig4}b, resulting in a characteristic HOM peak \cite{PRA2014High-Visibility}. The dots are the obtained coincidence counts, and the solid lines are the Gaussian fitting curves obtained by the Monte Carlo method\cite{molmer1993monte}, in which 1000-time random sampling is performed around the measured data. From the HOM peak, we obtain an FWHM of 1355.0$\pm$212.8 fs and a visibility of 47.87$\pm$4.32\%, limited by the non-perfect coupling ratio of the integrated photonic-plasmonic device. As a reference, we measure the photons-photons HOM interference using a 50:50 polarization-maintaining fiber coupler (PMC) in place of the integrated photonic-plasmonic device, observing an FWHM of 1143.8$\pm$37.9 fs and a visibility of 46.31$\pm$1.26\%. The time-resolved coincidence histogram is shown in Fig.~\ref{fig:Fig4}d, and the resulting interference peak is presented in Fig.~\ref{fig:Fig4}e.

\section{Conclusions}\label{sec3}
We conclude that our experiment demonstrates unambiguous quantum interference between photons and polaritons using an Au–SiN\textsubscript{x} integrated photonic-plasmonic device.~The observed visibility of 47.87$\pm$4.32\%, obtained using weak coherent-state single-photon wavepackets, confirms the preservation of quantum characteristics in the light-matter interaction process for the excitation of SPP. This measured visibility approaches the theoretical limit of 50\% for weak-coherent single-photon wavepackets, and could be further enhanced to ~95.74\% with genuine single-photon wavepackets\cite{fan2021effect}, offering even stronger validation of quantum coherence in photon–plasmon conversion processes.~The slight reduction in visibility and broadening of the interference envelope are attributed to fabrication-induced coupling ratio asymmetries and dispersion in the integrated photonic-plasmonic device, respectively.~Our results establish that SPPs can inherit the indistinguishability and coherence property of their excitation photons, confirming that quantum states are preserved throughout the light–matter interaction process.~This positions SPPs not only as passive intermediaries in quantum photonic systems, but also as real carriers of quantum information.~The involvement of free electrons in SPPs provides an excellent degree of freedom, enabling quantum-state control through external electric fields. The future integration of electrodes with photonic–plasmonic circuits could thus enable dynamic manipulation of the quantum information encoded in SPPs, offering a new and scalable approach to on-chip quantum information processing.~Our work thus lays the foundation for harnessing integrated photonic–plasmonic systems not only to investigate quantum coherence and entanglement at the nanoscale, but also to engineer and actively control quantum states using electronic means, thus opening a new pathway for quantum technologies by leveraging the unique light–matter interaction inherent in plasmonic systems.


\subsection*{Disclosures}
The authors declare no competing interests.

\subsection* {Data and Materials Availability} 
Data and materials are available upon request from the corresponding author.

\subsection* {Acknowledgments}
This work was supported by Quantum Science and Technology-National Science and Technology Major Project (Nos.~2024ZD0300800, 2021ZD0300701), Sichuan Science and Technology Program (Nos.~2024YFHZ0370, 2025YFHZ0339, 2024YFHZ0369,\linebreak 2024YFHZ0368), National Natural Science Foundation of China (Nos.~62475039, 62405046, 62375043), Tianfu Jiangxi Laboratory (No.~TFJX-ZD-2025-005).



\bibliography{nsr}   
\bibliographystyle{spiejour}   



\vspace{1ex}


\end{spacing}
\end{document}